%% file: sediment.tex
\documentstyle[12pt,twoside, epsf, here, psfig, natbib]{article}
\input{defs}

\title{Some Wave--related, Heavy Mineral Placer Deposits}

\author{S. J. Childs \\ {\small\em Department of Pure and Applied
Mathematics, Rhodes University, Grahamstown,} \\ {\small\em 6140, South
Africa} \\ \\ F.A. Shillington \\ {\small\em Department of Physical Oceanography, University of Cape Town, Rondebosch,} \\ {\small\em 7701, South Africa}}

\date{}       

\begin{document}

\maketitle

\begin{abstract}
\noindent {\em Examples of heavy mineral placer deposits
are presented in which wave reflection, refraction,
diffraction and resonance would appear to have played a
major concentrating role. Their geometry is compared with
the computer generated patterns predicted for the
reflection, refraction and diffraction of surface waves
moving over fairly simple, idealised bathymetries. Much of
this work is founded on the idea that similar sediments
document equivalent (or once equivalent) flow--tractional
environments. \\

\noindent Most of the examples could be satisfactorily explained in
this fashion. It may therefore be possible to ignore the exact
physics of the boundary layer, longshore and tidal return currents
etc. at the scale on which these examples occur, leaving the way open
for the qualitative use of results obtained using the likes of the
mild slope wave equation. A ``Monte Carlo'' approach based on wave
induced tractions should therefore succeed in elucidating presently
known heavy mineral placer deposits and, consequently, in predicting
other deposits which remain as yet undiscovered.}
\end{abstract}


\section{Introduction} \label{39}

Examples of heavy mineral placer deposits in which wave reflection,
refraction, diffraction and resonance would appear to have played a
major concentrating role, and in which it is believed a ``Monte Carlo''
approach based on wave induced tractions would work, are presented in
this paper. Some general factors thought to be instrumental in the
formation of heavy mineral placer deposits are listed, and a brief
description is given of the oceanographical context in which examples
occur off the west coast of Southern Africa.

The description of each example includes sonographs, a contour map, and
some possible formational mechanisms. Relevant diagrams, detail or
geological interpretation is also sometimes given. Much of this work
is founded on the idea that similar sediments document equivalent (or
once equivalent) flow--tractional environments.

Sorted sediments often form distinctive patterns on the sea bed. These
sometimes take the form of radiating gravel lenses around islands or
other positive topographic features. Other patterns are sometimes
associated with the leeward side of these positive topographic
features. Successive gravel lenses in parallel formation may occur at
the base of submarine cliffs. Many variations and combinations of these
themes do, however, also occur. Sonographs of examples, in which a
difference in sediment type is visible, appear in Section \ref{186}.
These features are discernible due to a pronounced difference in the
acoustic properties of the sediments. A significantly larger number of
intricate sediment distribution patterns are not visible.

Heavy mineral placers are well documented (eg. {\sc Hallam} (1989),
and {\sc Murray et al.} (1970)) examples and of particular economic
interest. Phosphatic gravels, metamorphic garnet, diamond,
kimberlitic garnet and sometimes even ilmenite, all rocks and
minerals of similar hydrodynamic character\footnotemark[1]
\footnotetext[1]{the total hydrodynamic effect of the sediment
grain's combined density, volume, shape/surface area etc.} but from a
diverse range of source environments are consistently associated. A
microcosm of hydrodynamic zonation can sometimes even be recognised
within these sediments. For example flatter, higher surface area
macles generally experience greater tractions and are consequently
often more amenable to longshore transport. (This oversimplification
is, of course, not always applicable as mobilisation is also
dependent on the reactive angle, bed roughness etc.)

The degree of hydrodynamic sorting, sometimes enrichment, and yet the
diverse environments of origin represented in these sediments,
undoubtedly suggest they were concentrated by physical processes. An
apparent correlation between these accumulations and regions where one
could expect to find energetically similar wave conditions, appears to
exist. For instance, the radiating gravel lenses coincide with a given
wave refraction pattern in the vicinity of islands. The successive
parallel lenses are often an integer multiple of an expected wavelength
apart, suggesting reflection and resonance off submarine cliffs. The
heavy mineral placers are often associated with caustics of a certain
energy.

Such interference probably occurs everywhere but is only apparent in
sonographs where there is an accoustic difference in sediment type. In
rather special instances, where the traction in caustics etc. exceeds
the threshold required to mobilise the overlying, fine sediment sheet,
yet is not sufficient to strip coarser, underlying sediments and
thereby expose the bedrock, these phenomena become visible to the naked
eye in sonographs. The accretion of bands of coarse terrigenous
sediment at a spacing one half the wavelength of the prevailing wave
climate is yet another phenomenon frequently observed.

The quality of the descriptive material unfortunately varies and some
of the formational mechanisms proposed are somewhat speculative,
consequently requiring a certain amount of patience and vision from the
reader.

\section{Factors Possibly Contributing to the Formation of Heavy
Mineral Placer Deposits}

Many of the phenomena presently listed can be explained in terms of a
modification to the onshore--offshore traction profile (two such
profiles are depicted in Fig. \ref{133} on page \pageref{133}). All
established wave theories predict an inverse correlation between
traction on the sea bed and depth up until the point of breaking.
This, somewhat simplistic, deduction is based on the components of
the stress tensor, calculated for each theory according to the
constitutive relation for a Newtonian fluid. It is on this basis that
the ``cartoon'' traction profile (featured twice in Fig. \ref{133})
is proposed. The profile suggests the possibility of twin
concentrations, one onshore and more deposition related, the other
lagging offshore, more extraction related. The onshore grouping
together of higher surface area macles which are more amenable to
transport\footnotemark[2], separate from offshore, highly
concentrated gem deposits would be an example of such twin deposits.
\footnotetext[2]{Traction is measured in units of force per unit
area.}

\subsection{The Significance of Marine Transgression}

The wave induced, shoreward motion of sediment (remembering that
orbitals are not perfectly closed in reality) is in the same
direction as the migration of the shoreline during transgressive
periods. Marine transgressions can therefore be expected to collect
and sort sediment from a considerable area before finally
concentrating it. By a converse argument regressive marine sequences
might be expected to have a relatively distributive effect on shelf
sediments.

Marine transgressions are characterised by thin, discontinuous
stratigraphic units and erosional surfaces. It is significant that
sorted sediment concentrations, especially heavy mineral placers, arise
as a result of the selective removal of particles of a certain
hydrodynamic character as well as the selective deposition of others.

\subsection{Gently Sloping Bathymetry}

A gently offshore--sloping bathymetry makes a far greater area
susceptible to an erosional range of tractions, hence a greater source
area is susceptible to extraction.

\subsection{Wave Converging Bathymetry}

Wave convergent bathymetry focuses and confines material extracted from
a larger than usual source area. Refraction also results in a far
greater range and zonation of wave energy thereby concentrating a
greater range of sediments in specific hydrodynamic zones.

\subsection{Interference Phenomena in General}

Refraction, diffraction, reflection, caustic interference,
resonance and the interaction with longshore and rip
currents are all potential mechanisms which favour the
concentration of heavy minerals. They also result in a far
greater range and zonation of wave energy, concentrating a
greater range of sediments in specific hydrodynamic zones.
Refraction is possibly most significant as a sediment
concentrating process.

\subsection{A ``Half Wavelength'' Effect} \label{401}

Bands of coarse terrigenous sediment regularly accrete at a spacing
half the wavelength of the prevailing wave climate. Bands of the type
described can be seen in Fig. 10 in the caustic zone
associated with the submerged canyon head (on the north--western side
of it) as well as in the vicinity of the other fragmented canyon head
(just south--west of Deposit 2). This phenomenon is easily understood
if one considers that the horizontal orbital velocity vanishes twice
within one wavelength. The particle path and streamline diagrams for a
sinusoidal progressive wave in Fig. \ref{193} (after {\sc Kinsman}
(1965)) illustrate well what is described. Only the sweeps and outward
interactions are thought to be significant in the bedload transport of
coarse terrigenous material (see the accoustic measurements of {\sc
Williams et al.} (1989)) and so--called ``half
wavelength'' phenomena are thought to form consequently.

Naturally, one would anticipate the proposed phenomenon to occur in the
context of standing waves, resonance etc. only. Statistical properties
of the near--bottom velocity field would otherwise not vary in the
along--propagation direction. The ``half wavelength'' phenomena are
indeed usually associated with bathymmetries suggestive of standing
waves and resonance. An alternative formational mechanism is a change
in the harmonic induced by the propagation of waves over an abrupt,
downward step. The orientation of the bands relative to the prevailing
swell direction suggests, however, that this is not the case. Some kind
of Bragg effect can also not be totally ruled out.

This process may well have been operative in other of the examples in
Section \ref{186} (in which case they will have been mistakenly
attributed to waves of a far shorter wavelength than those supposed in
terms of this ``half wavelength'' effect).

\subsection{Bed Roughness}

Potholes, gulleys, benches etc. can be significant in trapping heavy
minerals in a dynamic, high energy state equilibrium. The recognition
of bed roughness as an independent factor is in some ways a scale
distinction, although turbulence and other boundary layer processes
are almost certainly more significant in such instances. The finite
element simulation of the motion of a rigid body in a fluid with free
surface described ({\sc Childs} and {\sc Reddy} (1998)) is better
suited to dealing with such complications, although the fluid model
still needs to be adapted for turbulence.

\section{Sedimentation on the West Coast of Southern Africa}

Sea levels on the west coast of Southern Africa have fluctuated over a
vertical range in excess of one kilometre in recent geological times
({\sc Kennett} (1982)). Fairly long periods in which the sea level
remained constant are documented by submarine cliffs, drowned beaches,
ancient wave cut platforms and various other relict features (eg. the
cliffs which occur around sixty metres below the present sea level, but
many other examples exist, both above and below present sea level).
Lesser stillstands are documented by less significant ``notches''. The
cliffs are occasionally interrupted by canyons which mark the path of
ancient watercourses. It is these ancient watercourses which introduced
the heavy minerals to then low--lying areas, in what is now the sea.
The ancient watercourses and canyons therefore demarcate the original
source areas where terrigenous sediment was introduced to the sea. They
also give rise to unusual topographic features which are conducive of
wave refraction, diffraction, reflection and resonance.

The ancient river terraces were subsequently reworked by the sea so
that the terrigenous sediments have come to lie on the last surface
of unconformity in their present state. These sediments are often
presently found trapped against submarine cliffs (or lesser notches)
and have been re--concentrated by wave related phenomena (eg.
refraction, reflection, diffraction and resonance) associated with
the submarine canyons and other topographic features in the nearby
vicinity. Modern day beaches and other sediment formations indicate a
predominantly south westerly incident swell direction (although {\sc
de Decker} (1988), suggests a more southerly direction -- probably
the case in deeper water before refraction begins) and the coarse
terrigenous sediments are normally found concentrated on the north
eastern to northern side of positive topographic features.

The sediments on the west coast of Southern Africa form a typically
fining--upward transgressive sequence ({\sc Birch et al.} (1991)).
Present--day shelf sediments may be divided into two main facies for
the purposes of interpreting the side scan sonographs on pages to
follow.

{\bf Facies 1:} Fine, marine fallout consisting mainly of pelagic ooze,
shells, marine detritus etc. Due to its absorbent nature and ooze--like
consistency this mud is not a good reflector of sound and consequently
appears white in sonographs. This facies forms a highly mobile blanket
overlying much of the bottom and is mobilised during storm periods
(this fact is evident from side--scan sonographs of the same area taken
before and after storms).

{\bf Facies 2:} Coarse megarippled sediments (frequently hidden by
facies 1) are better reflectors of sound due to the larger, more
definite and harder surfaces of constituent grains. They are
characterised by megaripples which are usually of the order of one to
two metres apart and have heights in the region of one metre. They
frequently lend a finger--print like appearance to the coarse
terrigenous sediment features in sonographs eg. Figs. 7 and
6.

{\bf Facies 3:} Underlying bedrock is the most reflective and casts
jagged, white shadows. This normally lends an enhanced
three--dimensional effect to bedrock in sonographs.

Frequent scours through fine, overlying, marine detrital sediments
demarcate zones of higher traction on the sea bed, exposing underlying,
coarse, terrigenous sediments, alternatively bedrock. The
characteristic shapes of these scours suggest many were formed in
caustic zones. One might imagine that such interference occurs far more
frequently than the instances discernable in sonographs, the instances
in which there is a difference in the accoustic properties of the
sediments.

\section{Examples of Wave--related Placer Deposits} \label{186}

Sonographs of four areas in which sediments have formed distinctive
patterns on the sea bed follow. Relevant topography and a proposed
mechanism of formation is included in the discussion of these areas.
Special detail and geological interpretation is provided in some
cases.

\subsubsection*{Interpreting Sonographs}

Sonographs are ``pictures'' based on the intensity of echos, or
reflected sound. Highly reflective surfaces appear dark and absorbant
material or ``shadows'' appear white. In particular, sidescan
sonographs are derived from a repetitive series of scans, taken along
narrow beams perpendicular to the path of a combined source/receiver,
as it is towed above the area of interest. The sonographs presented
here are collages of those sidescan tracks.

The first rule to interpreting what results might therefore be to
regard the sonograph as the negative of a floodlit terrain
illuminated by rows of floodlights. In practice the tow fish is never
successfully towed at a constant height above the sea bed, or along a
perfectly straight line. The magnification, overlap and intensity of
individual tracks within the sidescan sonar collages consequently
varies and the tracks often required to severe buckling and smoothing
over when being glued down. The technology required to edit and
arrange sonographs, maps and diagrams was, unfortunately, also far
from ideal at the time of collection.

\subsection{Deposit 1 (Figs. 3, 5, 6 and
7)}

The sonograph in Fig. 3 shows fine, marine fallout sediments
to have been removed by waves, alternatively, deposition has been
prevented, concentrating the coarser, terrigenous gravel in parallel
zones of high traction near the base of a submarine cliff. The
associated bathymetry appears in Fig. 5. These particular
zones of high traction are believed to be the result of reflection
and refraction, off and around submarine cliffs. A schematic overlay
outlining the processes presently proposed is given in Fig.
4.

{\bf Reflection:} Inferred wave reflection off the three steepest
regions of the cliff (clearly visible in Fig. 5) is
particularly strong, giving rise to the crescentic gravel patches
visible in Fig. 3. Reflection off the remainder of the north
west facing slope is not as pronounced and gives rise only to long,
thin, parallel streamers. Careful examination of the sonograph Figure
3 reveals a total of four parallel gravel lenses orientated
perpendicular to the north west, although one is not as pronounced as
the others. The distance between the parallel lenses averages
approximately one hundred and twenty three metres. If it is assumed
that the bottom features are caused by reflected surface waves which
have the same wavelength as the spacing between the gravel features,
the dispersion relation for small amplitude Airy waves in water of
intermediate depth\footnotemark[1], \footnotetext[1]{Intermediate depth
is defined as a depth deeper than one twentieth, but shallower than
half a wavelength -- {\sc Le M\'{e}haut\'{e}} (1976)}
\[
\lambda = \frac{g T^2}{2 \pi} \tanh \left( \frac{2 \pi d}{\lambda}
\right),
\] 
can be used to determine the period, where $\lambda$ is the wavelength,
$T$ the period, $d$ the depth and $g$ the gravitational constant.
\begin{eqnarray*}
\lambda \approx 123m \Rightarrow T &\approx& \sqrt[2]{\frac{2 \pi \times 123}{9.81
\times \tanh(\frac{2 \pi 50}{123})}} \\
&=& 8.93 \mbox{ seconds.}
\end{eqnarray*}
It is significant that typical average wave periods for
this area are quoted as being a similar value ({\sc de
Decker} (1988)). It is therefore reasonable to expect that
parts of the gravel outline could visibly be seen to have
been modified where successive scans overlap given that the
rough scale analysis is correct. One such definite
alteration to the sediment was discernible in successive
overlapping tracks, suggesting that sedimentation was
active at the time the sonographs were taken.

{\bf Refraction:} The entire set of parallel lenses is
thought to be contained within a zone of high wave energy,
or modulated envelope, attributed to the refraction of
surface waves about topography lying to the south of that
depicted (c.f. the shaded regions in Figure 14). The
bottom traction due to incoming waves interfering with
reflected waves would be enhanced as it entered the high
energy zone attributed to refraction. The entire modulated
envelope containing the reflected gravel feature (or at
very least, that part occurring in deeper water) is
considered a typical refraction feature associated with
submarine cliffs to the south of the area depicted. Kinks
and necking in some parts of the gravel feature bear a
resemblance to those in the shaded regions in the
refraction feature in Fig. 14.

{\bf Mega-ripples:} Striations visible within the coarse terrigenous
material (see Figs. 6 and 7) are megaripples
with an amplitude of approximately one metre and a wavelength of
between one and two metres apart. They are a characteristic feature of
the coarse terrigenous sediments and form perpendicular to the
prevailing wave direction.

The orientation of the megaripples within the gravels, the
orientation of the entire reflection--related feature (both long
streamers and crecentic patches) and the orientation of the proposed,
refraction--related, modulated envelope relative to the cliffs, are all
consistent with a prevailing south westerly swell direction.

{\bf Alternative Formational Processes:} Current schools of thought
pertaining to the orientation of megaripples oppose the idea that the
feature is the result of resonance and caustics formed by waves
propagating over the cliffs from the south (assuming the megaripples
were not superimposed by a different wave condition). A formational
mechanism whereby waves are reflected simplistically into shallower
water would entail an incident wave angle which is improbable and there
would also then be no obvious explanation why the feature does not
repeat itself closer to the base of the submarine cliff.

\subsection{Deposits 2 and 3 (Figs. 8, 9 and
10)}

In Figs. 8 and 9 fine, marine fallout sediments
and silt would once again appear to have been removed, alternatively,
deposition prevented, in wave--induced zones of high traction, leaving
a ``window'' into coarser, terrigenous material. A ``cartoon'' of the
associated bathymetry and geological interpretation for both can be
found in Fig. 10 (the gravel features are the two darkly
shaded, inset patches).

{\bf Refraction:} The zones of higher traction are attributed to
refracted waves which originate from the south. These waves are
refracted about the canyon heads clearly evident in Fig. 10.
Reflection and resonance off the submarine cliffs and canyons are
processes, almost certainly, also operative. The three, uniform, gravel
lenses constitute a further feature to the left of the outcrop in the
upper part of the sonograph Fig. 9. They are considered a
typical refraction feature when associated with the outcrop of bedrock
to their right (compare with the refraction around a circular pile
depicted in Fig. 14). Such refraction related phenomena
about positive topographic features are typical and numerous examples
exist.

{\bf Reflection and Resonance:} The distance between the bands
averages roughly sixty metres \footnotemark[1] \footnotetext[1]{By
far the most common spacing between such gravel features.} in Fig.
8 and 30 metres in Fig. 9. Assuming some kind of
resonance or reflection \footnotemark[2] \footnotetext[2]{Unlikely to
be a straightforward reflection for both features since their
environments are in every other way alike, highlighting the need for
the algorithms to be proposed in Section \ref{180}.} off the adjacent
cliffs and canyons in the vicinity, one might guess $\lambda = 2
\times 60m$ in Fig. 8 (supposing the ``half wavelength''
effect) and $\lambda = 4 \times 30m$ in Fig. 9 (the integer
multiple, 4, is chosen for agreement alone and can only be explained
in a heuristic sense). Using this rough scale analysis in the
deep\footnotemark[3] \footnotetext[3]{Deep water is defined as water
of a depth exceeding half a wavelength -- {\sc Le M\'{e}haut\'{e}}
(1976)} water, Airy formula,
\[
\lambda = \frac{g}{2 \pi} T^2,
\] 
which implies
\begin{eqnarray*}
\begin{array}{cclcccl} T &=& \sqrt[2]{\displaystyle \frac{2 \pi \times
60 \times 2}{9.81}} & \hspace{10mm} \mbox{and} \hspace{10mm} & T &=&
\sqrt[2]{\displaystyle \frac{2 \pi \times 30 \times 4}{9.81}} \\
&\approx& 8.8 \mbox{ seconds} & & &\approx& 8.8 \mbox{ seconds}
\end{array}
\end{eqnarray*}
for the surface wave period expected to generate the bottom features in
Figs. 8 and 9. A period of 6.2 seconds (which
would correspond to $\lambda = 60m$) would, however, still lie just
within the limits of acceptability. The high traction zones, arising
from the refraction of such waves, would probably only be capable of
mobilising the fine, overlying sediment sheet.

{\bf ``Half Wavelength'' Effect:} In Fig. 10, in the caustic
zone associated with the submerged canyon head (on the north--western
side), as well as in the vicinity of the other fragmented canyon
head (just south--west of Deposit 2), fields of coarse terrigenous
sediment, which have accreted at a spacing half the wavelength of the
prevailing wave climate, occur. In the latter instance this may be a
bunching up of traction zones, much the same as that evident for the
island and circular pile in Fig. 14, or a superposition of
phenomena arising from more than one, alternatively a more complex,
topographic feature. In the former instance, however, resonance of the
type described on page \pageref{401} would appear to be the only
likelihood. 

Detailed maps giving exact geological information based on samples and
seismic profiles could unfortunately not be printed due to the
economically sensitive nature of the deposits.

\subsection{Deposit 4 (Figs. 11, 12, 13 and
15)}

The radiating geometry of the sediment within and about the submerged
archipelagos highlighted in Figs. 12 and 11 is
remarkably reminiscent of the surface wave refraction patterns in
Fig. 14. Presently in deep water of the order of one hundred
metres on the shelf, the relative bathymetry of underlying and
surrounding bedrock is given in Figs. 13 and 15 (the
$x + \cdot$ quantities denote relative depth). 

The recent sediments are visible in the side--scan sonograph and
overlay, Figs. 11 and 12. Zones of higher traction are
believed to arise due to the combined effects of refraction and
diffraction of waves originating from the south (see Fig. 14
-- all wave theories predict a relationship between traction on the
sea bed and absolute wave height).

The distance between the terrigenous bands varies from anything between
three hundred to four hundred and fifty metres. Using the dispersion
relation for small amplitude Airy waves in water of intermediate depth
\[
\lambda = \frac{g T^2}{2 \pi} \tanh \left( \frac{2 \pi d}{\lambda}
\right)
\] 
\begin{eqnarray*}
\Rightarrow T &\approx& 13 \mbox{ to } 16 \mbox{ seconds.}
\end{eqnarray*}

This observation that sediments in Deposit 4 are only sorted by the
effects of abnormally large waves is consistent with the depths at
which the sediments occur.

The accompanying geological sample maps of this area are extremely
interesting although they could unfortunately not be included. Highly
sorted sediments can be found concentrated in island--induced
caustics which would correspond to a relict sealevel.

It is possible that refractive processes operative on a far greater
scale than those to which the deposits have presently been attributed,
were responsible for the formation of some of these examples,
highlighting the need for the algorithms proposed in Section \ref{180}.
What does appear to be apparent, however, is that the physical
complexity of viscous effects etc. in the boundary layer may be
qualitatively ignored at the scales on which these features occur.

\section{A ``Monte Carlo'' Approach to Sorted Sedimentation in a Wave
Environment} \label{180}

Finite element algorithms based on an inputed incident wave height,
angle and wavelength are able to predict relevant fields, such as
velocity and wave height, or related quantities which are dependent on
wave reflection, refraction and diffraction over large areas when run
on computers which could be said to have fairly average capabilities by
today's standards. Algorithms which calculate the traction (surface
force) on the sea bed can, in turn, be written based on these fields.

The algorithms are written on the premise of laminar flow and other
gross simplifications which circumvent the usual problems associated
with modelling the exact physics of the problem, as might well have
been expected. The relevant, outputed fields may therefore not suffice
where the more intricate mechanics of the boundary layer are concerned
and the fields may require modification according to ``velocity
defect'' and other laws ({\sc Tennekes} and {\sc Lumley} (1972)). The
obvious geometry of the examples discussed in the previous section and
the scale on which these deposits occur suggest, however, that the need
for modification is unlikely in calculations of the qualitative type
about to be proposed.

Although obvious uses for algorithms which calculate tractions on the
sea bed exist for known wave climates and sea levels, it is often
aperiodic, catastrophic events which are of relevance in sedimentology.
In some cases such events may not even occur as often as once in a
lifetime.  The known wave climate is therefore not necessarily the
applicable one and it is therefore also likely that many field studies
observing wave--sediment associations are meaningless.

This section proposes a way in which to elucidate the problem of sorted
sediments and their environments of formation, keeping the
afore--mentioned facts in mind and with an awareness that a whole range
of conditions may contribute to the formation of some intricate
distribution of sorted sediment. It is proposed that the distinctive
patterns described in this paper may be likened to unique signatures
which could be deciphered with the use of a little lateral thinking. If
one could establish the exact characteristics of the wave environment
operative at the time of formation, regions where similar tractions,
and hopefully the same sediments, arise could be located elsewhere. A
method to do just that is what this section concerns itself with.

Much of what is about to be proposed is founded on the idea that the
hydrodynamic properties of individual sediment particles document the
conditions under which the sediment formed. Sediments said to share the
same hydrodynamic character are those supposed to have been
concentrated under similar flow--tractional conditions. ``Sorted'', as
used here, refers to sediments in which a common hydrodynamic character
is obviously apparent. Sorted sediment concentrations arise as a result
of the operation of two simultaneous processes viz. the selective
deposition of particles of a certain hydrodynamic character and the
selective removal (erosion) of those of a different hydrodynamic
character. Using tractions, this section attempts to understand sorted
sediments from a more empirical point of view. Much of this work is
founded on the idea that similar sediments document equivalent, or once
equivalent, flow tractional environments.

\subsection{The Underlying Logic to a ``Monte Carlo'' Approach}

Supposing the hydrodynamic character of sediments to document the
conditions under which they formed, it is proposed that the correct
sediment--concentrating wave condition/s may be deduced by employing
the logic
\begin{eqnarray*}
\begin{array}{lll} \mbox{\it an incorrect wave condition} & \Rightarrow
& \mbox{\it no necessary correlation in bed} \\ & & \mbox{\it
flow-tractional environments},
\end{array}
\end{eqnarray*}
therefore,
\begin{eqnarray*}
\begin{array}{lll}
\mbox{\it a necessary correlation in bed} & & \\ \mbox{\it
flow-tractional environments} & \Rightarrow & \mbox{\it the correct
wave condition}.
\end{array}
\end{eqnarray*}
This is, of course, assuming that an equal or greater number of sorted
sediment concentrations (distinctive ``pattern features'') arise than
the number of parameters required to specify a prevailing wave climate
or condition; furthermore, that the traction acting on these sediments
always varies with deep water wave specification. By ``wave
conditions'' is also meant sea level, parameters of bed roughness and
any other, suspected controlling variables in addition to the more
obvious incident wave angle, wave height and wave period. A correlation
in tractions at positions where similar sediment concentrations occur
would be looked for using different wave conditions.

\subsection{Using the Algorithms as a Qualitative as Opposed to a
Quantitative Tool.}

A number of reasons exist for using algorithms as a qualitative rather
than quantitative tool. The mild slope wave equation is one obvious
candidate on which to base such large scale calculations which forgo
the exact physics of the boundary layer. The fact that qualitative
capabilities of the mild slope wave equation are in question where
slopes greater than 1 : 3 ({\sc Booij} (1983)) are involved, boundary
effects, bed roughness (the mild slope wave equation assumes the sea
bed to be locally flat), particle--particle interactions and the
unknown final static or dynamic equilibrium state of deposits are all
factors which may contribute to making exact quantitative calculations
meaningless.

The comparison of relative traction values would obviously have to be
kept within reason. For example, a condition which results in a wave
base far above the sea bed would produce a perfect correlation in
tractions (zero traction) on the sea bed.

\section{Conclusions and Further Research}

Side--scan sonar measurements show interesting, dune--like terrigenous
sediment features on the sea floor. Agreement between these features
and zones of high traction predicted by crude refraction, reflection
and diffraction studies would appear to exist. These studies suggest
that the physical complexity of viscous effects etc. in the boundary
layer may be ignored at the scales on which these features occur. Very
crude scale analyses show that the ``wavelength'' of these features
corresponds to surface waves with realistic periods.

It would appear that a sediment's hydrodynamic character and its
geographical distribution may be likened to a finger--print. In cases
this ``finger--print'' may indicate the prevailing wave climate. This
idea that heavy minerals lie locked beneath the waves by a
oceanographical ``combination lock'' (whose code is incident wave
angle, wave height, wave period and sea level) may seem slightly
fanciful, however, such concentrations do seem to exist. It is
conceivable that adjustments to the coastline in presently active
wave environments could be made to concentrate heavy minerals.

The author is presently in the process of completing a finite element
algorithm which calculates tractions on the sea bed, based on a
finite element solution of the mild slope wave equation, and which
uses a biquadratic least squares fit to model the sea bed locally in
the vicinity of a node. This will be used in an attempt to determine
the conditions under which the sediments depicted in the sonographs
were concentrated.

\section{Acknowledgements}

Benco and De Beers Marine are thanked for the sonographs and
associated bathymetries. Jacobus Williams, Timothy Gregory and Shaun
Courtney are thanked for assistance in photographing, scanning,
editing and arranging the material in a presentable format. The use
of George Ellis' computer is gratefully acknowledged, and Kevin
Collville is also thanked in this regard. 


\nocite{b:3}
\nocite{b:2}
\nocite{birch:1}
\nocite{b:1}
\nocite{me:2}
\nocite{dedecker:1}
\nocite{dias:1}
\nocite{g:1}
\nocite{hallam:1}
\nocite{hockney:1}
\nocite{kennett:1}
\nocite{kinsman:1}
\nocite{lh:1}
\nocite{l:2}
\nocite{dbm:1}
\nocite{p:2}
\nocite{rosenthal:1}
\nocite{frank:1}
\nocite{tennekes:1}
\nocite{williams:1}
\nocite{heathershaw:1}
\nocite{heathershaw:2}
\nocite{zienk:1}

\bibliography{sediment}

\newpage
\section{Captions}

\begin{figure}[H] 
\begin{center} \leavevmode
\mbox{\epsfbox{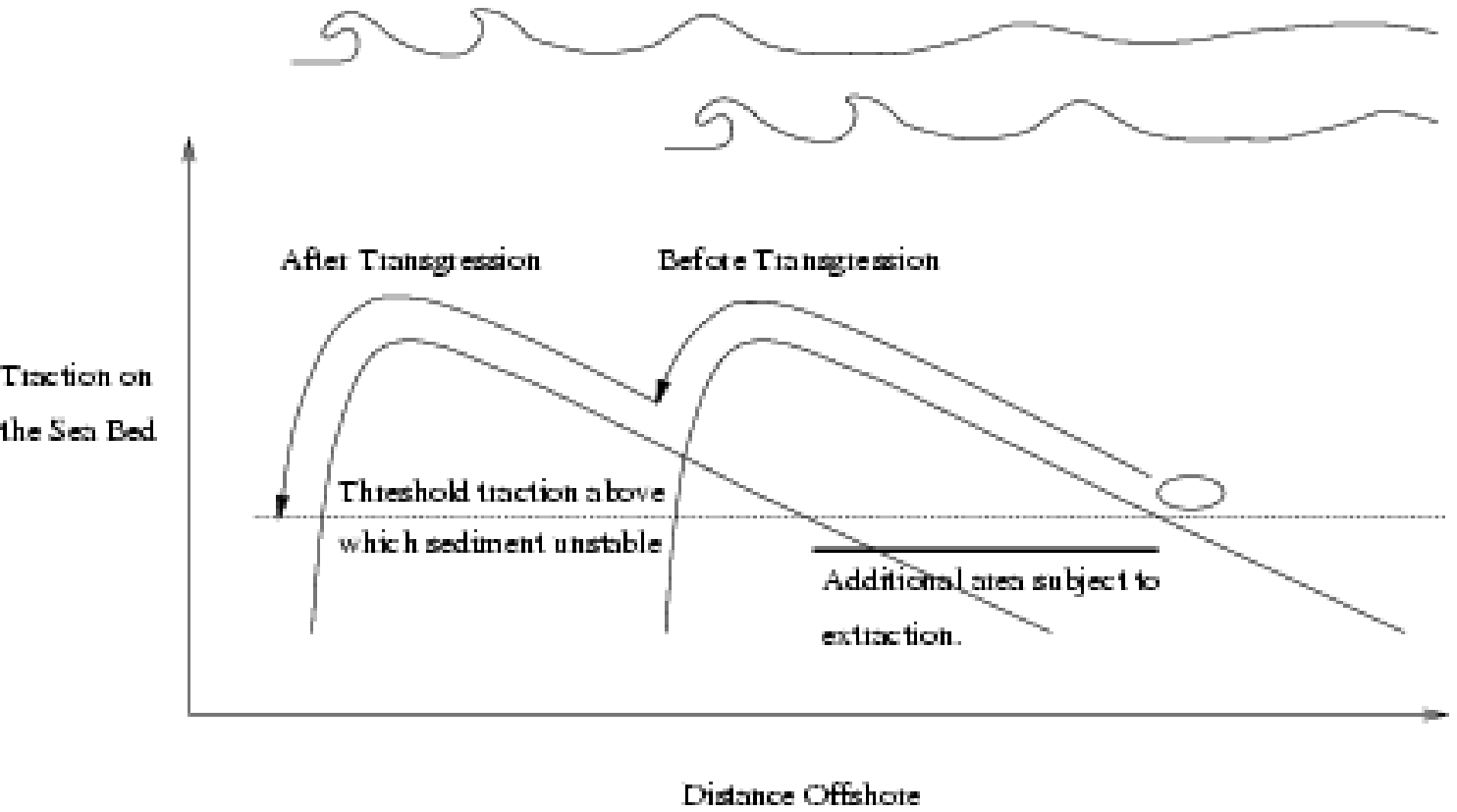}}
\end{center}
\caption{} \label{133} 
\end{figure}

Fig. \ref{133}: A ``Cartoon'' Profile of Tractions on the Sea Bed
Before and After a Marine Transgression Illustrating the Sediment
Collecting Effect

\begin{figure}[H] 
\begin{center} \leavevmode
\mbox{\epsfbox{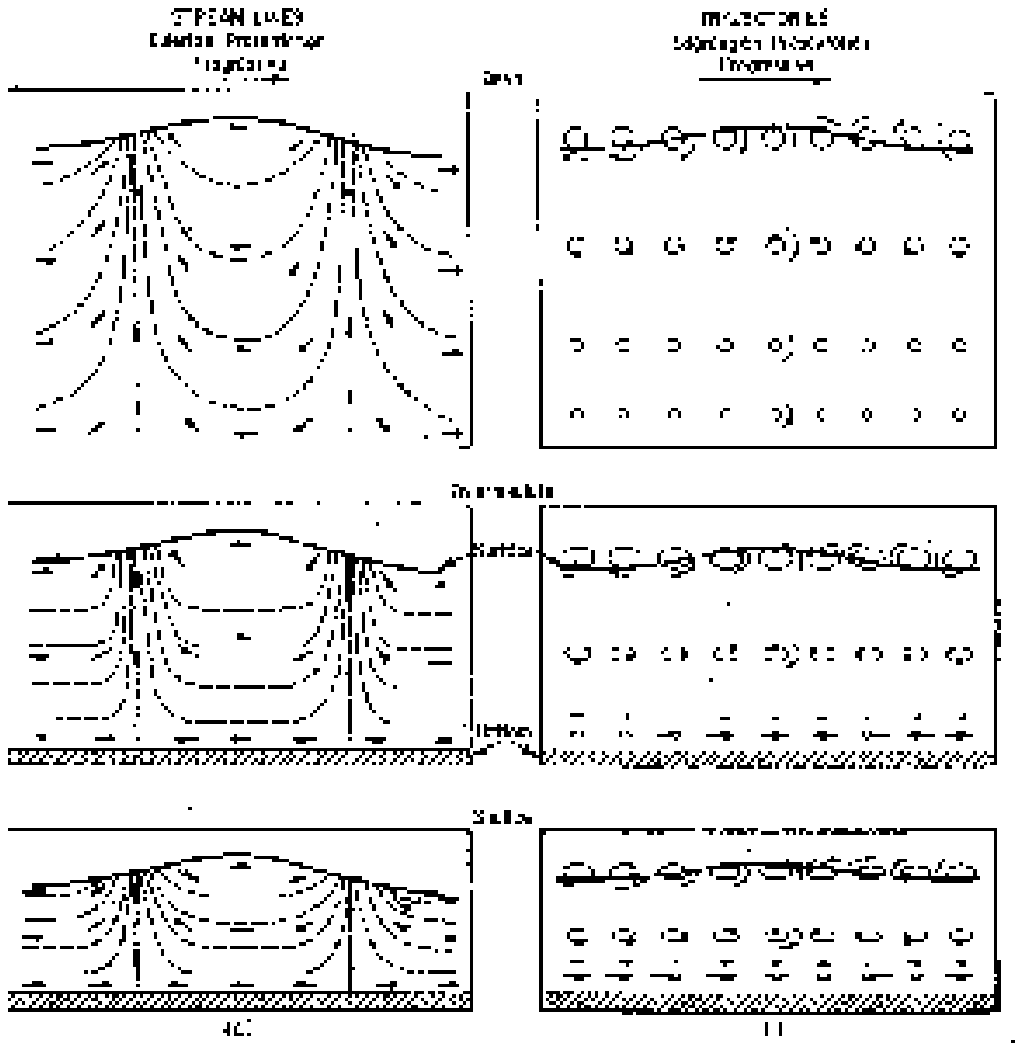}}
\end{center}
\caption{} \label{193} 
\end{figure}

Fig. \ref{193}: Particle Path and Streamline Diagrams for a
Sinusoidal Progressive Wave (after {\sc Kinsman} (1965))

Fig. 3: Side--scan Sonograph of Deposit 1

Fig. 4: Overlay Schematic Diagram of the Proposed
Wave Processes which Gave Rise to the Gravel Formations in
Deposit 1

Fig. 5: The bathymetry in and around Deposit 1 (See Figure
3).  An outline of the feature is lightly sketched. The contour
interval is 1$m$ and the grid interval 1$km$.

Fig. 6: An Enlargement of Part of Deposit 1 (the Feature
Depicted in Fig. 3)

Fig. 7: An Enlargement of Part of Deposit 1 (the Feature
Depicted in Fig. 3)

Fig. 8: Side--scan Sonograph of Deposit 2 

Fig. 9: Side--scan Sonograph of Deposit 3 

Fig. 10: The Bathymetry and Geological Interpretation of
Deposits 2 and 3 (grid interval = 1$km$)

Fig. 11: Side--scan Sonograph of Deposit 4 

Fig. 12: Overlay to Fig. 11

Fig. 13: The Relative Bathymetry in Deposit 4 (contour
interval = 1$m$).

Fig. 14: Relative, absolute wave heights associated with a
single island (after {\sc Pos et al.} (1987)) and
relative, absolute wave heights associated with a circular pile (after
{\sc Berkhoff} (1976)). Note that these are not wave fronts but instead
``modular envelopes''. All wave theories predict a relationship between
traction on the sea bed and absolute wave height.

Fig. 15: The Relative Bathymetry In and Around Deposit 4
(Fig. 13 bathymetry inset). contour interval = 1$m$

\newpage

\end{document}

%% file: defs.tex
\newcounter{tony}
\newcommand{\beq}{\begin{equation}}
\newcommand{\eeq}[1]{\label{eq:#1}\end{equation}}

\newcommand{\eqref}[1]{(\ref{eq:#1})}

      \newcommand{\beqn}{\begin{equation}}
      \newcommand{\eeqn}{\end{equation}}
      \newcommand{\beqna}{\begin{eqnarray}}
      \newcommand{\eeqna}{\end{eqnarray}}

%% file: sediment.bbl
\begin{thebibliography}{}

\bibitem[Berkhoff(1976)]{b:3}
Berkhoff, J. C.~W. (1976).
\newblock Mathematical models for simple harmonic linear water waves wave
  diffraction and refraction.
\newblock Technical report, Delft hydraulics laboratory.

\bibitem[Bettess and Zienkiewicz(1977)]{b:2}
Bettess, P. and Zienkiewicz, O.~C. (1977).
\newblock Diffraction and refraction of surface waves using finite and infinite
  elements.
\newblock {\em International Journal For Numerical Methods In Engineering},
  {\bf 11}, 1271--1290.

\bibitem[Birch {\em et~al.}(1991)]{birch:1}
Birch, G.~F., Day, R.~W., and Plesis, A.~D. (1991).
\newblock Nearshore quaternary sediments on the west coast of southern africa.
\newblock Technical report, Republic of South Africa Department of Mineral and
  Energy Affairs.

\bibitem[Booij(1983)]{b:1}
Booij, N. (1983).
\newblock A note on the accuracy of the mild--slope wave equation.
\newblock {\em Coastal Engineering}, {\bf 7}, 191--203.

\bibitem[Childs and Reddy(1998)]{me:2}
Childs, S.~J. and Reddy, B.~D. (1998).
\newblock Finite element simulation of the motion of a rigid body in a fluid
  with free surface.
\newblock {\em accepted, Computer Methods in Applied Mechanics and
  Engineering}.

\bibitem[de~Decker(1988)]{dedecker:1}
de~Decker, R.~H. (1988).
\newblock The wave regime on the inner shelf south of the orange river and its
  implications for sediment transport.
\newblock {\em South African Journal of Geology}, {\bf 91(3)}.

\bibitem[Dias and Iooss(1993)]{dias:1}
Dias, F. and Iooss, G. (1993).
\newblock Capillary--gravity solitary waves with damped oscillations.
\newblock {\em Physica D}, {\bf 65}, 399--423.

\bibitem[Gonsalves(1985)]{g:1}
Gonsalves, J.~W. (1985).
\newblock Estuary mouth stability project progress report no. 2 modifications
  to the finite element programme `{W}ave'.
\newblock Technical Report T/SEA 8515, Council for Scientific and Industrial
  Research.

\bibitem[Hallam(1989)]{hallam:1}
Hallam, C.~D. (1989).
\newblock {\em The Geology of Coastal Deposits of Southern Africa.}, volume~2
  of {\em Ore Deposits of South Africa}.

\bibitem[Hockney(1985)]{hockney:1}
Hockney, A.~P. (1985).
\newblock Suggested pleistocene climatic changes and their possible effects on
  the raised and submerged beaches of the {W}est {C}oast of {S}outhern
  {A}frica.
\newblock Technical report, De Beers Marine.

\bibitem[Kennett(1982)]{kennett:1}
Kennett, J.~P. (1982).
\newblock {\em Marine Geology}.
\newblock Prentice--Hall.

\bibitem[Kinsman(1965)]{kinsman:1}
Kinsman, B. (1965).
\newblock {\em Wind Waves, their Generation and Propagation on the Ocean
  Surface}.
\newblock Applied Mathematical Sciences. Prentice--Hall.

\bibitem[Longuet-Higgins and Stewart(1964)]{lh:1}
Longuet-Higgins, M.~S. and Stewart, R.~W. (1964).
\newblock Radiation stresses in water waves; a physical discussion, with
  applications.
\newblock {\em Deep--Sea Research}, {\bf 11}, 529--562.

\bibitem[M\'{e}haut\'{e}(1976)]{l:2}
M\'{e}haut\'{e}, B.~L. (1976).
\newblock {\em An Introduction to Hydrodynamics and Water Waves}.
\newblock Springer--Verlag, New York.

\bibitem[Murray {\em et~al.}(1970)]{dbm:1}
Murray, L.~G., Joynt, R.~H., O'Shea, D.~C., Foster, R.~W., and Kleinjan, L.
  (1970).
\newblock The geological environment of some deposits off the coast of {S}outh
  {W}est {A}frica.
\newblock {\em Institute of Geological Sciences Report}, {\bf 70}(15),
  119--141.

\bibitem[Pos {\em et~al.}(1987)]{p:2}
Pos, J.~D., Kilner, F.~A., and Fischer, P.~G. (1987).
\newblock Combined refraction--diffraction of water waves by an island.
\newblock {\em International Journal of Engineering Science}, {\bf 25}(5),
  577--590.

\bibitem[Rosenthal(19)]{rosenthal:1}
Rosenthal, G.~N. (19).
\newblock {\em Lift Forces on Spherical Particles Near a Horizontal Bed in
  Oscillatory Flow}.
\newblock Ph.D. thesis, University of Cape Town.

\bibitem[Shillington and Britten-Jones(1979)]{frank:1}
Shillington, F.~A. and Britten-Jones, A. (1979).
\newblock Features of surface waves off the {S}outhern {C}ape coast and their
  associated meteorological conditions during a sever storm between 30 {A}ugust
  and 3 {S}eptember 1978.
\newblock {\em South African Journal of Science}, {\bf 75}.

\bibitem[Tennekes and Lumley(1972)]{tennekes:1}
Tennekes, H. and Lumley, J.~L. (1972).
\newblock {\em A First Course in Turbulence}.
\newblock {MIT} press.

\bibitem[Williams and Caldwell(1988)]{williams:1}
Williams, A.~T. and Caldwell, N.~E. (1988).
\newblock Particle size and shape in pebble beach sedimentation.
\newblock {\em Marine Geology}, {\bf 82}, 199--215.

\bibitem[Williams {\em et~al.}(1989a)]{heathershaw:2}
Williams, J.~J., Thorne, P.~D., and Heathershaw, A.~D. (1989a).
\newblock In situ acoustic measurements of marine gravel threshold and
  transport.
\newblock {\em Sedimentology}, {\bf 36}(1), 61--74.

\bibitem[Williams {\em et~al.}(1989b)]{heathershaw:1}
Williams, J.~J., Thorne, P.~D., and Heathershaw, A.~D. (1989b).
\newblock Measurements of turbulence in the benthic boundary layer over a
  gravel bed.
\newblock {\em Sedimentology}, {\bf 36}(6), 959--971.

\bibitem[Zienkiewicz and Heinrich(1979)]{zienk:1}
Zienkiewicz, O.~C. and Heinrich, J.~C. (1979).
\newblock A unified treatment of steady--state shallow water and
  two--dimensional {N}avier--{S}tokes equations -- finite element penalty
  function approach.
\newblock {\em Computer Methods in Applied Mechanics and Engineering}, {\bf
  17/18}, 673--698.

\end{thebibliography}
